# Prediction of intrinsic topological superconductivity in Mn-doped GeTe monolayer from first-principles

Xiaoming Zhang,[1,2,3] Kyung-Hwan Jin,[4,3] Jiahao Mao,[5] Mingwen Zhao,[6] Zheng Liu,[2*] and Feng Liu[3*]

[1] *Department of Physics, College of Information Science and Engineering, Ocean University of China, Qingdao, Shandong 266100, China*
[2] *Institute for Advanced Study, Tsinghua University, Beijing 100084, China*
[3] *Department of Materials Science and Engineering, University of Utah, Salt Lake City, Utah 84112, USA*
[4] *Department of Physics, Pohang University of Science and Technology, Pohang 37673, Republic of Korea*
[5] *State Key Laboratory of Low Dimensional Quantum Physics and Department of Physics, Tsinghua University, Beijing, 100084, China*
[6] *School of Physics and State Key Laboratory of Crystal Materials, Shandong University, Jinan, Shandong 250100, China*

*Correspondence to: fliu@eng.utah.edu, zheng-liu@mail.tsinghua.edu.cn

**Recent discovery of topological superconductors (TSCs) has sparked enormous interest. Realization of TSC requires a delicate tuning of multiple microscopic parameters, which remains a great challenge. Here, we develop a first-principles approach to quantify realistic conditions of TSC by solving self-consistently Bogoliubov-de Gennes equation based on Wannier function construction of band structure, in presence of Rashba spin-orbit coupling, Zeeman splitting and electron-phonon coupling. We further demonstrate the power of this new method by predicting the Mn-doped GeTe ($Ge_{1-x}Mn_xTe$) monolayer – a well-known dilute magnetic semiconductor showing superconductivity under hole doping – to be a Class D TSC with Chern number of -1 and chiral Majorana edge modes. By constructing a first-principles phase diagram in the parameter space of temperature and Mn concentration, we propose the TSC phase can be induced at a lower-limit transition temperature of ~40 mK and the Mn concentration of $x$~0.015%. Our approach can be generally applied to TSCs with a phonon-mediated pairing, providing useful guidance for future experiments.**

# INTRODUCTION

The topological phase of superconductors (SC) has recently received intense research interest as the superconducting quasi-particles residing in the non-trivial gapless/zero-energy boundary states are considered a form of Majorana fermions. Majorana fermions are their own anti-particles [1] and obey the non-Abelian exchange statistics [2], which can be utilized for topological quantum computation [3]. Topological superconductors (TSC) exhibit various exotic phenomena, including zero modes on the magnetic vortex [4], "fractional" Josephson effect [5], non-local correlation [6], and thermal responses [7]. By now, the theoretical aspects of TSCs are reasonably well understood, but the experimental confirmation remains a great challenge due to the requirement of tuning multiple microscopic parameters like the Fermi level, magnetic field, temperature, etc.. Hence, it is highly desirable to predict more TSCs and quantify experimental conditions to advance the field.

Unlike the successful first-principles prediction of electronic and topological materials, theoretical predictions of TSCs are challenging because of the uncertainty in the parameters used to construct Bogoliubov-de Gennes (BdG) Hamiltonian. Usually, only the pre-conditions of TSC, e.g. Rashba splitting [8] or topological properties [9-12] in the normal state of known superconductors, were analyzed using first-principles method, but not the topology of superconducting quasi-particles. Instead, effective models of TSC states are constructed with empirical parameters, at the best partially fit to the first-principles results [13]. Moreover, conventional first-principles approaches that estimate the superconducting transition temperature ($T_c$) by employing the empirically McMillan's formula [14] or solving the Migdal-Eliashberg formula [15] cannot be applied to the cases involving spin-orbit coupling (SOC) and magnetism (internal or external). Therefore, more versatile and accurate methods to predict $T_c$ for SC as well as TSC are highly desirable.

In this article, we attempt to further extend first-principles calculations to the field of TSCs, by developing a verstile approach to quantify realistic conditions of TSC. We construct and solve self-consistently a material-specific first-principles BdG Hamiltonian, based on Wannier function construction of band structure, in presence of Rashba SOC, Zeeman splitting and electron-phonon coupling (EPC). Furthermore, we demonstrate the usefulness of this new method by predicting the Mn-doped GeTe ($Ge_{1-x}Mn_xTe$) monolayer to be a TSC by constructing a first-principles phase diagram in the parameter space of temperature and Mn concentration.

Generally, TSC materials can be classified as intrinsic or extrinsic, depending on the experimental conditions of realizing the non-trivial phase. Intrinsic TSCs exhibit inherently a non-trivial superconducting gap without the need of applying an external field or constructing a heterostructure. They may be *p*-wave SCs with natural spin-triplet pairing [16,17], such as $Sr_2RuO_4$ [18], Cu/Sr/Nb-doped $Bi_2Se_3$ [19] and non-centrosymmetric SCs [20], or *s*-wave SCs with an effective spin-triplet pairing resulting from helical spin-polarized states, such as the two-dimensional (2D) topological electronic states [21,22], and 1D [23,24] and 2D Rashba electronic states [25-27] which belong to the so-called Class D TSC without time-reversal symmetry (TRS). Extrinsic TSCs employ the same physical mechanisms, but realization of their non-trivial properties requires applying external fields or constructing heterojunctions. To the best of our knowledge, all the known Class D TSCs formed by the *s*-wave superconductivity are extrinsic, such as the semiconductor nanowire with strong SOC [28], the ferromagnetic atomic chains [29], the nanoscale magnetic islands [30], the ferromagnet [31], and the topological surface [32] and edge states [33] proximitized with conventional SCs with/without applying external magnetic field. Notably, the signature of TSCs observed by applying an external magnetic field in a superconducting material, e.g., $FeTe_{0.55}Se_{0.45}$ [34], epitaxial GeTe [35] and *β*-$Bi_2Pd$ thin film [36] indicates the possible existence of intrinsic Class D TSC without needing the external magnetic field, which will further enrich the physics of TSC, in the same perspective as from quantum Hall effect (with magnetic field) to anomalous quantum Hall effect (without).

Given the necessary conditions for realizing Class D TSCs with 2D Rashba electrons [25-27], i.e. inversion symmetry breaking, Zeeman gap opening and superconductivity, the IV-VI compound GeTe with Mn doping, a dilute magnetic semiconductors (DMS) with a ferromagnetic Curie temperatures $T_c^{FM}$ up to ~200 K for epitaxial layers on $BaF_2$ (111) substrate [37-42], caught our attention. The superconductivity of GeTe with *p*-type doping due to Ge vacancy was confirmed as early as the 1960s [43,44]. It is also known as a ferroelectric material with rhombohedral layered, non-centrosymmetric structure below the ferroelectric Curie temperature of ~700 K [45]. Recently, a gradual opening of Zeeman gap in the Rashba bands of GeTe with Mn doping was observed, attributed to the entanglement of ferromagnetic and ferroelectric order [46]. Also, a recent experiment has reported possible signatures of extrinsic TSC in GeTe film under external magnetic field [35].

Specifically, we focus on the recent experimentally exfoliated GeTe monolayer [47], which was predicted to be useful in optoelectronic devices and may be a type-II Ising superconductor upon slight hole doping [48,49]. We first show that GeTe monolayer inherits all the key characteristics of its bulk phase by using conventional first-principles calculation. Then, the first-principles BdG Hamiltonian was constructed via a Wannier-function scheme, through which we found that the GeTe monolayer with the hole concentration of ~$7.4 \times 10^{13}$ cm$^{-2}$ becomes superconducting below ~120 mK and the $Ge_{1-x}Mn_xTe$ monolayer is a Class D TSC with $T_c$~40 mK characterized by a non-zero Chern number and chiral Majorana edge modes. A phase diagram of $Ge_{1-x}Mn_xTe$ is constructed by employing the newly developed first-principles approach to guide experimental detection of the predicted SC and TSC phase. Since both the exfoliated GeTe monolayer [47] and epitaxial $Ge_{1-x}Mn_xTe$ thin film already exist [37-42], our prediction should be readily testable experimentally. Our approach provides a benchmark to make material-specific predictions of TSCs by using first-principles calculations.

## RESULTS AND DISCUSSION

**Crystal and electronic band structure (conventional first-principles calculation)**

The crystal structure of GeTe monolayer is shown in Fig. 1a, which is a (111) layer fragment of its bulk phase. Each Ge(Te) atom is bonded with three Te(Ge) atoms, forming a buckled honeycomb lattice. The in-plane lattice constant $a$ and buckling height $h$ was optimized to be ~3.955 Å and ~1.565 Å, respectively, in good agreement with previous report [48]. Due to the absence of inversion symmetry, a large Rashba splitting arises in the electronic band structure (Fig. 1b). The electronic states are doubly degenerate at the Γ and M points, forming the so-called Kramers pairs, while the degeneracy was lifted away from these time-reversal invariant points. For the four valence bands near the Fermi level of our interest, hereafter we name the lower (upper) two bands as the Rashba (Ising) bands for clarity, referring to their respective electronic spin-texture near the Γ point (Supplementary Fig. 1).

To predict the TSC formed by 2D Rashba electrons [25-27], we focus on the Rashba bands with a significant Rashba splitting coefficient $\alpha_R$ = 0.66~0.76 eV Å. It is comparable with that of heavy metals Au(111) and Bi(111) surface [50,51], but slightly smaller than that of bulk GeTe [52]. A strong Rashba effect is desirable for the electrons to overcome the suppressing effect of Zeeman field on superconductivity. Doping 0.1 holes per primitive cell, corresponding to a hole concentration of ~7.4×10$^{13}$ cm$^{-2}$, will move the Fermi level ($E_F$) to the Dirac point formed by the Rashba splitting (Fig. 1b). The electronic density of states (DOS) at $E_F$ is thus increased from 0 to $N_F$~1.4 states/eV/primitive-cell, which stems mainly from the $p$-orbitals of Te and Ge atoms (Supplementary Fig. 2). Figure 1c shows the spin-texture $s(k)$ on the Fermi surfaces (FSs) of the 0.1-hole-doped GeTe monolayer. One can clearly see the in-plane components are helical-like, while the out-of-plane ones are small. The metallic nature and the anti-parallel spins at the **k** and **-k** points provide the prerequisite conditions for s-wave superconductivity.

Having demonstrated the Rashba spin splitting in the GeTe monolayer, we now discuss the second ingredient, the Zeeman gap. It has been reported that a Zeeman gap can be opened in the bulk Ge$_{1-x}$Mn$_x$Te with a ferromagnetic order parallel to the (111) direction [46], which is the easy magnetization direction for small $x$ [53]. By reproducing the experimental results of bulk Ge$_{1-x}$Mn$_x$Te based on the virtual crystal approximation (VCA) [54], the spin state of Mn dopants was determined to be $S$=5/2 (Supplementary Note 1). Consequently, the out-of-plane high-spin state ($S$=5/2) of Mn dopants is adopted in Ge$_{1-x}$Mn$_x$Te monolayer under VCA. As expected, the Zeeman gaps $\delta_z$ of Rashba and Ising bands opened at Γ increase monotonically with the increasing Mn concentration (Fig. 1d), and can be fit by the equation of $\delta_z$ = 250×$x$ and $\delta_z$ = 1550×$x$ meV, respectively. The different slopes are resulted from different out-of-plane spin component of the electronic states near the Dirac point versus near the valance band maximum (VBM) (Supplementary Fig. 1c).

**Superconductivity (newly developped first-principles approach)**

We next discuss the phonon mediated superconductivity of the 0.1-hole-doped GeTe monolayer. From the calculated phonon spectra (Fig. 2a), we first confirm its dynamical stability by the absence of imaginary frequency. For the acoustic branch with the lowest vibration frequency, Kohn anomalies can be seen at certain **q**-points around Γ, which is favorable for enhancing EPC. The **q**- and $v$-resolved EPC $\lambda_{\mathbf{q}v}$ show two significant features (Fig. 2a). On one hand, all phonon modes can couple with electrons. This is further confirmed by the comparison between the phonon DOS $F(\omega)$ and the isotropic Eliashberg spectral function $\alpha^2 F(\omega)$ (Fig. 2b). To estimate the relative contributions of acoustic and optical modes, we calculated the cumulative EPC $\lambda(\omega) = 2\int_0^\omega d\omega' \frac{\alpha^2 F(\omega')}{\omega'}$. The $\lambda(\omega)$ increases quickly to 1.13 at the frequency of ~10 meV, which is about 81% of the total EPC $\lambda$=1.39. This indicates that the EPC stems mainly from the acoustic modes. The convergence of EPC calculation has been carefully checked (Supplementary Note 2). On the other hand, only the vibration modes with a finite wave vector can couple with electrons. This is because for all the FS contours surrounding Γ (Fig. 1c), only a finite length of phonon wave vectors can connect the initial and final scattering states. Additionally, both $\alpha^2 F(\omega)$ and $\lambda_\mathbf{q}$ illustrate that the soft modes associated with the Kohn anomalies help to enhance EPC [55].

To estimate the superconducting transition temperature ($T_c$), we construct a material-specific BdG Hamiltonian $H_{BdG}(\mathbf{k})$ by employing the electronic Hamiltonian $H_{MLWFs}(\mathbf{k})$:

$$H_{BdG}(\mathbf{k}) = \begin{pmatrix} H_{MLWFs}(\mathbf{k}) - E_F & \\ & -H^*_{MLWFs}(-\mathbf{k}) + E_F \end{pmatrix} + H_\Delta \quad (1)$$

$$H_\Delta = \Delta \left( \varphi^\dagger_{i\uparrow} \varphi^\dagger_{(i+\frac{\mathbf{x}}{2})\downarrow} - \varphi^\dagger_{(i+\frac{\mathbf{x}}{2})\downarrow} \varphi^\dagger_{i\uparrow} \right) + \Delta \left( \varphi_{(i+\frac{\mathbf{x}}{2})\downarrow} \varphi_{i\uparrow} - \varphi_{i\uparrow} \varphi_{(i+\frac{\mathbf{x}}{2})\downarrow} \right) \quad (2)$$

Here the basis of $H_{MLWFs}(\mathbf{k})$ is the Maximally localized Wannier functions (MLWFs) and obtained by fitting the first-principles band structure. The chemical potential $E_F$ in $H_{BdG}(\mathbf{k})$ is the Fermi energy where the superconducting gap condenses. Then we formulate the gap equation into the following form:

$$\Delta_{mn} = -\frac{g_{mn}}{2V} \sum_{l,\mathbf{k}>0} \frac{1}{1+e^{\beta \varepsilon_{l,\mathbf{k}}}} \frac{\partial \varepsilon_{l,\mathbf{k}}}{\partial \Delta_{mn}} \quad (3)$$

Here the intra-orbitals spin-singlet pairing $H_\Delta$ ensures $m=n$. The absolute pairing strength $g_{nn}$ is assumed identical for each band [55] and calculated as $g_{nn} = (\lambda-\mu^*)/N_F$. The Coulomb repulsion $\mu^*$ is described by the Morel-Anderson pseudopotential $\mu^* = \frac{\mu_c}{1+\mu_c \ln(\varepsilon_F/\theta_D)}$ [56], where the $\varepsilon_F$ can be regarded as the energy difference between the Fermi level and band maximum of the Ising bands. Only the electronic states within one Debye temperature $\theta_D$ around Fermi level are summed over. Formulating this gap equation enables us to solve the superconducting gap self-consistently at different temperature based on the eigenvalues $\varepsilon_{l,\mathbf{k}}$ of the constructed $H_{BdG}(\mathbf{k})$.

We emphasize that this new method is not only different from the conventional method employing the McMillan's formula [14] or solving the anisotropic Migdal-Eliashberg formula [15] in estimating $T_c$, but also extend the first-principles approach to calculate the topological invariant of superconducting gap and the critical magnetic field/doping-concentration of superconductivity (see below). We check the correctness of Eq. 3 by reducing it to the well-known gap equation of $1 = \frac{g}{V} \sum_{\mathbf{k}>0} \frac{1}{\sqrt{\Delta^2+E_\mathbf{k}^2}} \tanh\left(\frac{\beta}{2}\sqrt{\Delta^2+E_\mathbf{k}^2}\right)$ for a single-band s-wave SC [57]. Its reliability is further confirmed by reproducing superconductivity of three representative known SCs, i.e. bulk lead (Supplementary Fig. 4) [58], bulk GeTe (Supplementary Fig. 3d) [44], and MoS$_2$ monolayer (Supplementary Fig. 5) [59,60]. In the calcualtions, the Debye temperature $\theta_D$ of ~105, ~200, and ~300 K is employed for lead, GeTe, and MoS$_2$ monolayer [61,62], respectively. The weak Coulomb screening effect was considered by using $\mu^*$=0.2 for hole-doped bulk GeTe with 2.1×10$^{21}$ holes/cm$^3$ [44] and the electron-electron correlation was taken into account by reducing EPC by ~45.5% for electron-doped MoS$_2$ monolayer with 1.2×10$^{14}$ electrons/cm$^{-2}$ [63]. Details can be seen form Supplementary Note 3.1.

For the 0.1-hole-doped GeTe monolayer, we assume the Debye temperature (~200 K) and Coulomb repulsion $\mu^*$ to be same as that of bulk GeTe and extract the MLWFs using the $p$ orbitals of Ge and Te. Also, we heuristically reduce the calculated λ from 1.39 to ~0.76 by ~45.5%, based on the benchmark of correlation effect in MoS$_2$ monolayer [63]. This should set a lower limit on EPC since the correlation effect of $p$-orbitals is usually weaker than that of $d$-orbitals. The resulting $g_{nn}$ (~0.4) is comparable to that of bulk GeTe (~0.49) [64], which enables us to predict the superconducting gap $\Delta$ of the 0.1-hole-doped GeTe monolayer at different temperatures. From Fig. 2c, one can see the calculated $\Delta$ ~18.6 μeV for both Rashba and Ising bands, which is gradually suppressed with the increasing temperature. The $T_c$ is around ~120 mK, lower than that of GeTe film [44]. We anticipate that the predicted 2D superconductivity may be confirmed by growing GeTe monolayer on Si(111) wafers, as the epitaxial GeTe thin film was observed to be superconducting on this substrate [35].

Next we simulate the superconductivity of Mn-doped GeTe (Ge$_{1-x}$Mn$_x$Te) monolayer by adding an out-of-plane Zeeman energy $B_z$ in $H_{MLWFs}(\mathbf{k})$ first:

$$H^z_{MLWFs}(\mathbf{k}) = B_z \sigma_z \otimes I_{\frac{\aleph}{2} \times \frac{\aleph}{2}} + H_{MLWFs}(\mathbf{k}) \quad (4)$$

Here $\sigma_z$ is the Pauli matrix in spin space and the $I_{\frac{\aleph}{2} \times \frac{\aleph}{2}}$ is a $\frac{\aleph}{2} \times \frac{\aleph}{2}$ identity matrix. Then the new BdG Hamiltonian $H^z_{BdG}(\mathbf{k})$ can be constructed through the Eq. 1 and Eq. 2. The reliability of such treatment in simulating the SC without TRS is confirmed by reproducing the in-plane critical magnetic field of MoS$_2$ monolayer (Supplementary Note 3.2) [65].

By diagonalizing the $H^z_{MLWFs}(\mathbf{k})$ with different $B_z$ in momentum-space, we obtain the Zeeman gap $\delta'_z$ of Rashba and Ising bands opened at Γ point (Supplementary Fig. 6a), which can be fit as $\delta'_z = 0.122 \times B_z$ and $\delta'_z = 2.0 \times B_z$ meV, respectively. Combining with the δ$_z$ fit to the first-principles results in Fig. 1d, one obtains the relationship between $B_z$ and Mn concentration, as $B_z = 2049 \times x$ and $B_z = 775 \times x$ meV for the Rashba and Ising bands, respectively. The self-consistently calculated $T_c$ (Supplementary Fig. 6b) and $\Delta$ (Supplementary Fig. 6c) domenstrate that they both decrease gradually with the increasing $B_z$ due to the pairing breaking effect of magnetism. The superconductivity of Rashba (Ising) bands is fully superssed when $B_z > 0.35$ (0.23) meV, indicating a critical Mn doping concentration of $x_c$~0.017% (0.03%) (Fig. 2d). This value of $x_c$ is two orders of magnitude smaller than that (2%) of Mn doped MgB$_2$ [66], which is reasonable since the $T_c$ of GeTe monolayer is lower than MgB$_2$ by similar magnitude.

**Topological superconductivity and phase diagram (newly developed first-principles approach)**

To realize TSC formed by 2D Rashba electrons, model analysis proposes that the half of the Zeeman gap opened at the Dirac point of Rashba bands, i.e. $\delta_z/2$, should be larger than the superconducting gap $\Delta$ [25-27]. In the following, the first-principles approach has been extended to characterize the TSC phase based on a material-specific BdG Hamiltonian $H_{BdG}^z(\mathbf{k})$ for the first time. Specifically, we take $\Delta=0.2$ meV and $B_z=7.5$ meV with $\delta_z\sim0.9$ meV to construct the $H_{BdG}^z(\mathbf{k})$ of $Ge_{1-x}Mn_xTe$ monolayer via Eq. 1, Eq. 2 and Eq. 4. The relatively large $B_z$ and $\Delta$ are used to show the topological non-triviality more clearly. The $H_{BdG}^z(\mathbf{k})$ is analogous to the single-particle Hamiltonian of electrons with an energy gap mathematically. By diagonalizing $H_{BdG}^z(\mathbf{k})$ in the momentum space, we obtain the dispersion relation of superconducting quasi-particles (Fig. 3a). One can clearly see that the superconducting gap is indeed opened, where the topological invariant, i.e. first Chern number ($N_c$), is well-defined.

For 2D systems, the Chern number of $l$-th band is calculated by integrating the Berry curvature $\Omega^l(\mathbf{k}) = \nabla \times A^l(\mathbf{k})$ over the first Brillouin zone (BZ): $N_c^l = \frac{1}{2\pi}\int_{BZ}\Omega^l(\mathbf{k})d^2\mathbf{k}$, where $A^l(\mathbf{k}) = i\langle u^l(\mathbf{k})|\partial_{\mathbf{k}}u^l(\mathbf{k})\rangle$ is Berry connection. The total Chern number $N_c$ can be obtained by summing up the Chern numbers of all the states below the superconducting gap, which is quantized to -1. The Berry curvature resides mainly at the $\Gamma$ point associated with the Zeeman gap opening (Fig. 3b), similar to the band inversion in the quantum anomalous Hall systems. Here we should emphasize that $N_c$ does not physically correspond to a quantized Hall conductance because charge is not conserved in the BdG Hamiltonian [22]. Two chiral Majorana edge modes localized at two different edges clearly exist in the continuous superconducting gap due to the bulk-boundary correspondence (Fig.3c and 3d). The propagation of chiral Majorana fermions could lead to same unitary transformation as that in braiding Majorana zero modes [67], and the deterministic creation and braiding of chiral edge vortices in hybrid structures were elaborated [68].

We finally construct a phase diagram of the 0.1-hole-doped $Ge_{1-x}Mn_xTe$ monolayer in Fig. 4, to help guide future experimental detection of the predicted TSC phase formed by the superconducting Rashba bands. At the zero-temperature limit, the SC phase of the Rashba bands will be preserved for $x < x_c=0.017\%$ and the TSC phase will arise when $x > x_{min}=0.014\%$, where the pre-condition of $\delta_z/2>\Delta$ can be met (Supplementary Fig. 6d). At finite temperature, both the ferromagnetic and SC order should exist simultaneously for the formation of TSC phase. Referring to the ferromagnetic Curie temperatures $T_c^{FM}$ of $Ge_{1-x}Mn_xTe$ that increases linearly with increasing Mn concentration up to $x=0.2$ and can be fit by $T_c^{FM}(x)=333\times x$ K (Supplementary Fig. 7a) [37-42], we estimate $T_c^{FM}(x)$ and $x$-dependent $T_c$ will cross over at $x'_{min}$=0.014%= $x_{min}$ (Supplementary Fig. 7b), too. Consequently, the TSC phase can be formed for $x_{min} < x < x_c$ at the temperature where the SC order occurs. We suggest preparing the desired $Ge_{1-x}Mn_xTe$ monolayer on $BaF_2$ (111) substrate [37-42] by molecular beam epitaxy (MBE) since the growth is known to start in a 2D manner [39]. We anticipate that the chiral Majorana edge modes of $Ge_{1-x}Mn_xTe$ monolayer can be detected using Josephson effect [5] or charge transport [69], and controlled by magnetic flux [70]. The effects of magnetic anisotropy and GeTe film thickness on the TSC phase are discussed in Supplementary Note 4 and Note 5.

Lastly, in addition to the monolayer $Ge_{1-x}Mn_xTe$ we demonstrated here, we suggest three more candidate materials for Class D TSC. First, it was theoretically reported a ferromagnetic order can be induced by hole doping in monolayer GaSe, attributed to the exchange splitting of electronic states that exhibit a sharp van Hove singularity below the Fermi level [71]. Given the superconductivity of bulk GaSe under pressure and Ga/GaSe layers being made experimentally [72,73], the hole-doped monolayer GaSe is likely to be superconducting. Secondly, the magnetic order induced Rashba-Zeeman splitting in the Si-terminated surface of $HoRh_2Si_2$ has been observed experimentally [74]. A spin-singlet Cooper pairing could be tunneled into this surface state by superconducting proximity effect. Thirdly, since the heterostructures of $MnBi_2Te_4/Bi_2Te_3$ [75,76] and $Bi_2Te_3/NbSe_2$ [32,77] have already been fabricated, the $MnBi_2Te_4/Bi_2Te_3/NbSe_2$ hold high possibility to be synthesized. We propose that this type of heterostructures are also a Class D TSC characterized with non-zero Chern number. By applying our newly developed first-principles BdG Hamiltonian approach, a complete phase diagram of these systems can be constructed in the near future.

**METHODS**

**Details of the First-principles Calculations**

The Vienna *ab initio* simulation pack (VASP) [78,79] was utilized to calculate the electronic property of normal states based on the density functional theory (DFT). The exchange-correlation of electrons was treated within the generalized gradient approximation in the form of Perdew-Burke-Ernzerhof [80]. The atomic structures of GeTe monolayer and thin film was set up by introducing a vacuum region of more than 15 Å to avoid the interactions between neighboring images. Structural relaxations and self-consistent calculations as well as the Zeeman gap calculations were performed on a uniform 30×30×1 (18×18×18) **k**-point sampling of the first BZ for monolayer (bulk) GeTe. The energy cutoff was set to 400 eV for plane-wave basis. The dipole correction was used to cancel the artificial electric field imposed by the periodic boundary condition of GeTe thin film.

The QUANTUM ESPRESSO (QE) package [81] was used to calculate the phonon spectra and EPC based on the density-functional perturbation theory (DFPT) [82] as well as fit the first-principles band structure by interfacing the WANNIER90-2.1 code [83]. The Optimized Norm-Conserving Vanderbilt Pseudopotential [84] was employed and the kinetic energy cutoff was set to 100 Ry for wave functions. The hole doping was simulated by removing electrons from intrinsic GeTe monolayer and introducing the compensating jellium background to avoid divergence. The dynamic matrix and phonon frequency are computed on a 18×18×1 **q**-point mesh with a 18×18×1 **k**-point sampling, and a finer 36×36×1 **k**-point grid is used in the EPC calculations, where the DOS is converged (Supplementary Fig. 2b). Other **q/k**-point samplings (Supplementary Table 1) are also employed to check the convergence of EPC calculations. The phonon DOS $F(\omega)$ and the isotropic Eliashberg spectral function $\alpha^2F(\omega)$ as well as the cumulative frequency-dependent EPC $\lambda(\omega)$ are calculated using a 60×60×1 **q**-point sampling by means of the Fourier interpolation.

**Constructing the BdG Hamiltonian**

To perform first-principles prediction of TSC, the main challenge is to construct a BdG Hamiltonian of superconducting quasi-particles from the electronic Hamiltonian of normal state. Here we propose a strategy to overcome this obstacle by employing the MLWFs $\varphi_{MLWFs} = (\varphi_{i\uparrow}, \varphi_{(i+\frac{\aleph}{2})\downarrow})^T$ of specific materials. Here $\aleph$ is the total number of MLWFs with the orbital index of $i=1\ldots\frac{\aleph}{2}$ and $i+\frac{\aleph}{2}=\frac{\aleph}{2}+1\ldots\aleph$ for up and down spin orbitals, respectively. The starting point is the real-space Hamiltonian $H_{MLWFs}^{(\mathbf{R})}$ and the matrix element $h_{mn}^{(\mathbf{R})}$ represents the hopping between the *m*th and the *n*th MLWF connected by a vector of **R**, which can be obtained by fitting the first-principles band structures of specific materials using WANNIER90-2.1 code [83]. The matrix element $h_{mn}(\mathbf{k})$ of momentum-space Hamiltonian $H_{MLWFs}(\mathbf{k})$ of normal state is then obtained through $h_{mn}(\mathbf{k}) = \sum_{\mathbf{R}} h_{mn}^{(\mathbf{R})} \times e^{i\mathbf{k}\cdot\mathbf{R}}$. Within the mean-field approximation, we can write the material-specific BdG Hamiltonian $H_{BdG}(\mathbf{k})$ (see Eq. 1) under the basis vector of $\psi = (\varphi_{i\uparrow}, \varphi_{(i+\frac{\aleph}{2})\downarrow}, \varphi_{i\uparrow}^{\dagger}, \varphi_{(i+\frac{\aleph}{2})\downarrow}^{\dagger})^T$. The matrix element of $-H_{MLWFs}^{*}(-\mathbf{k})$ in $H_{BdG}(\mathbf{k})$ has the form of $h_{MN}(\mathbf{k}) = -h_{mn}^{*}(-\mathbf{k}) = -\sum_{\mathbf{R}} h_{mn}^{*(\mathbf{R})} \times e^{i\mathbf{k}\cdot\mathbf{R}}$ with $M = m + \aleph$ and $N = n + \aleph$. The constructed $H_{BdG}(\mathbf{k})$ is a $2\aleph \times 2\aleph$ matrix with the particle-hole symmetry $PH_{BdG}(\mathbf{k})P^{-1} = -H_{BdG}^{*}(-\mathbf{k})$, where $P = \tau^x \otimes I_{\aleph \times \aleph}$.

For the materials with external/internal magnetism, we first add a Zeeman term **B** in $H_{MLWFs}(\mathbf{k})$ using the vector of Pauli matrix **σ**:

$H^{\mathbf{B}}_{MLWFs}(\mathbf{k}) = \mathbf{B}\boldsymbol{\sigma} \otimes I_{\frac{N}{2}\times\frac{N}{2}} + H_{MLWFs}(\mathbf{k})$. Then the first-principles BdG Hamiltonian $H^{\mathbf{B}}_{BdG}(\mathbf{k})$ without TRS can be constructed for a specific material by using the above procedure.

**Formulating the gap equation**

Under the basis of $\Psi_{\mathbf{k}} = (c_{1,\mathbf{k}}, c_{2,\mathbf{k}}, ..., c^{\dagger}_{1,-\mathbf{k}}, c^{\dagger}_{2,-\mathbf{k}}, ...)^{\dagger}$, the multi-band Hamiltonian with s-wave pairing can be written as:

$$H = \sum_{\mathbf{k}>0} \Psi^{\dagger}_{\mathbf{k}} H_{BdG}(\mathbf{k}) \Psi_{\mathbf{k}} \text{ with } H_{BdG}(\mathbf{k}) = \begin{pmatrix} h(\mathbf{k}) & -\Delta \\ \Delta & -h^*(-\mathbf{k}) \end{pmatrix}$$

With the relation of $\frac{\partial H}{\partial \Delta_{mn}} = -\sum_{\mathbf{k}}(c^{\dagger}_{m,\mathbf{k}} c^{\dagger}_{n,-\mathbf{k}} + c_{n,-\mathbf{k}} c_{m,-\mathbf{k}})$, we can derive the gap equation of $\Delta_{mn}$ as:

$$\Delta_{mn} = \frac{g_{mn}}{V} \sum_{\mathbf{k}} \langle c_{n,-\mathbf{k}} c_{m,\mathbf{k}} \rangle = -\frac{g_{mn}}{2V} Tr\left[e^{-\beta H} \frac{\partial H}{\partial \Delta_{mn}}\right] / Tr\left[e^{-\beta H}\right]$$
$$= \frac{g_{mn}}{2\beta V} \frac{\partial}{\partial \Delta_{mn}} \ln Tr\left[e^{-\beta H}\right]$$

Here $m$ and $n$ are the band index. Diagonalizing the Hamiltonian $H$ into the form of $H = \sum_{\mathbf{k}>0,l} \varepsilon_{l,\mathbf{k}} d^{\dagger}_{l,\mathbf{k}} d_{l,\mathbf{k}}$, we have

$$\Delta_{mn} = \frac{g_{mn}}{2\beta V} \frac{\partial}{\partial \Delta_{mn}} \ln \sum_{\{n_{l,\mathbf{k}}=0,1\}} \left[e^{-\beta \sum_{l,\mathbf{k}>0} \varepsilon_{l,\mathbf{k}} n_{l,\mathbf{k}}}\right] = \frac{g_{mn}}{2\beta V} \frac{\partial}{\partial \Delta_{mn}} \sum_{l,\mathbf{k}>0} \ln\left(1 + e^{-\beta \varepsilon_{l,\mathbf{k}}}\right)$$
$$= -\frac{g_{mn}}{2V} \sum_{l,\mathbf{k}>0} \frac{1}{1+e^{\beta \varepsilon_{l,\mathbf{k}}}} \frac{\partial \varepsilon_{l,\mathbf{k}}}{\partial \Delta_{mn}}$$

Solving this gap equation self-consistently enables us to extimates the superconducting transation temperature $T_c$ and the critical magnetic field/doping-concentration of specific materils based on the material-specific BdG Hamiltonian $H_{BdG}(\mathbf{k})$ with/without TRS constructed from the state-of-art first-principles approach.

**ACKNOWLEDGEMENTS**

F.L. acknowledges financial support from DOE-BES (No. DE-FG02-04ER46148). X.Z. acknowledges financial support by the National Natural Science Foundation of China (No. 12004357) and the Young Talents Project at Ocean University of China (No. 862001013185). M.Z. acknowledge financial support from the National Natural Science Foundation of China (Nos. 21433006 and 11774201). K.J. acknowledges the support from Korea Research Fellowship Program through the National Research Foundation of Korea (NRF) funded by the Ministry of Science and ICT (Grant No. 2019H1D3A1A01071056).

**Figures:**

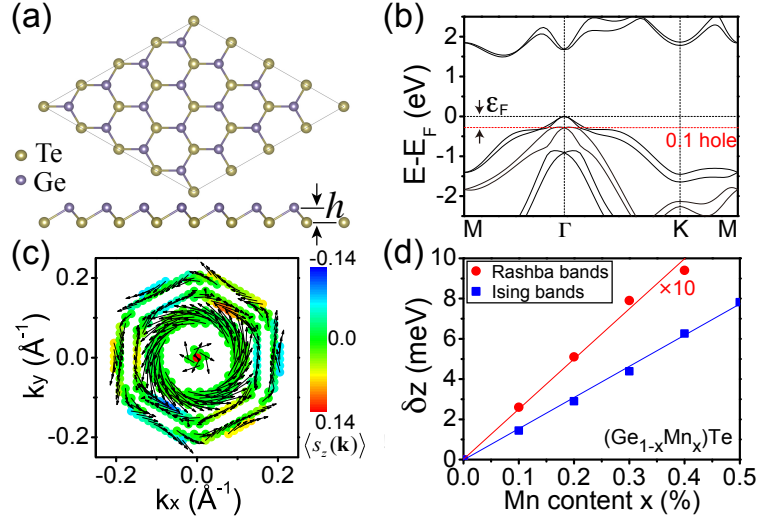

**Fig. 1. Crystal structure and electronic property.** (**a**) The top and side view of GeTe monolayer and (**b**) its electronic band structure. The black horizontal dashed line represents the Fermi level and the red one is the Fermi level after doping 0.1 holes per primitive cell, corresponding to the hole doping concentrations of ~$7.4\times10^{13}$ cm$^{-2}$. (**c**) The electronic spin-texture on the FS of 0.1-hole-doped GeTe monolayer. (**d**) The Zeeman gaps ($\delta_z$) of the Rashba/Ising bands opened at the $\Gamma$ point versus the concentration $x$ of Mn dopants with high-spin state. The $\delta_z$ of the Rashba bands are magnified by 10 times for clarity. The red and blue lines are the linearly fittings of $\delta_z$.

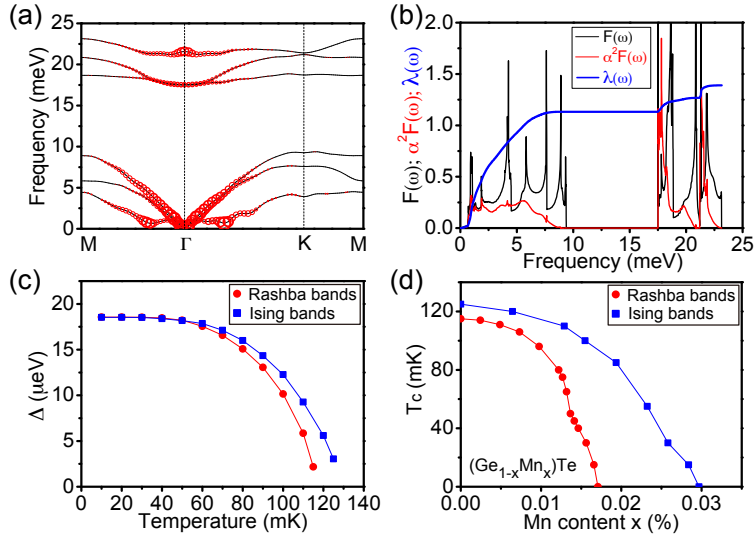

**Fig. 2. The superconductivity of 0.1-hole-doped GeTe monolayer without/with Mn dopants.** (**a**) Phonon spectra with the magnitude of EPC $\lambda_{\mathbf{q}\nu}$ being drawn proportional to the size of red circles. (**b**) The plots of phonon DOS $F(\omega)$, Eliashberg spectral function $\alpha^2 F(\omega)$, and cumulative frequency-dependent EPC $\lambda(\omega)$. (**c**) The temperature-dependent superconducting gap Δ and (**d**) the *x*-dependent translation temperature $T_c$ of the Rashba bands (red dots) and the Ising bands (blue squares) calculated on 600×600×1 **k**-points sampling in first BZ.

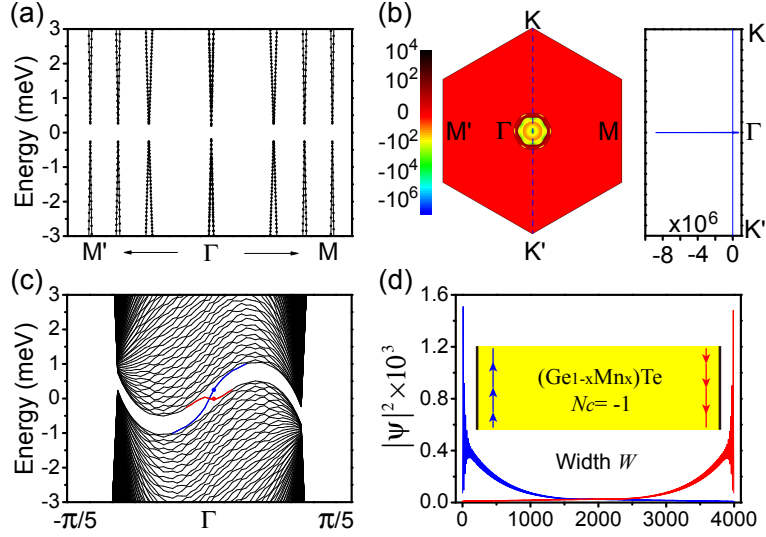

**Fig. 3. The topological superconductivity of 0.1-hole-doped Ge$_{1-x}$Mn$_x$Te monolayer.** (**a**) The dispersion relation of quasi-particles calculated by the first-principles BdG Hamiltonian. (**b**) The distribution of total Berry curvatures for the states below the superconducting gap in the first BZ (left panel) and along the line of K-Γ-K' (right panel). (**c**) The energy spectra of quasi-particles in a nanoribbon with the width (W) of 4000 primitive cells. The chiral Majorana edge modes are highlighted by red and blue lines. (**d**) The spatial distributed probability density $|\psi|^2$ of the in-gap states indicted by red and blue dot in (**c**). The inset is a schematic diagram of chiral Majorana edge modes with different propagation direction on opposite edges.

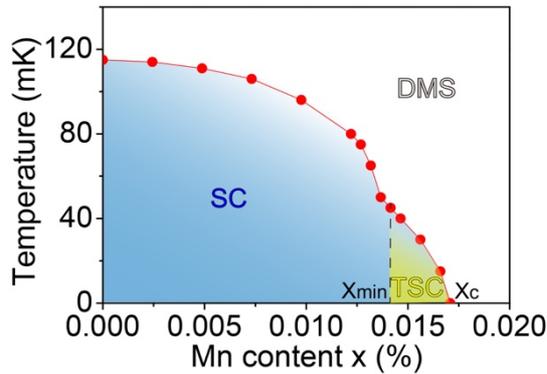

**Fig. 4. The schematic phase diagram of the 0.1-hole-doped Ge$_{1-x}$Mn$_x$Te monolayer.** The yellow shaded region marks the TSC phase.

Supplementary information for

# Prediction of intrinsic topological superconductivity in Mn-doped GeTe monolayer from first-principles

Xiaoming Zhang[1,2,3], Kyung-Hwan Jin[4,3], Jiahao Mao[5], Mingwen Zhao[6], Zheng Liu[2*], and Feng Liu[3*]

[1] *Department of Physics, College of Information Science and Engineering, Ocean University of China, Qingdao, Shandong 266100, China*
[2] *Institute for Advanced Study, Tsinghua University, Beijing 100084, China*
[3] *Department of Materials Science and Engineering, University of Utah, Salt Lake City, Utah 84112, USA*
[4] *Department of Physics, Pohang University of Science and Technology, Pohang 37673, Republic of Korea*
[5] *State Key Laboratory of Low Dimensional Quantum Physics and Department of Physics, Tsinghua University, Beijing, 100084, China*
[6] *School of Physics and State Key Laboratory of Crystal Materials, Shandong University, Jinan, Shandong 250100, China*

*Correspondence to: fliu@eng.utah.edu, zheng-liu@mail.tsinghua.edu.cn

**Supplementary Notes:**
Note 1. Determining the spin-state of Mn dopant
Note 2. The convergence of EPC λ
Note 3. Validating the newly developed first-principles approach
Note 4. The effect of magnetic anisotropy on the TSC phase
Note 5. The effect of GeTe film thickness on the TSC phase
**Supplementary Figures:**
Figure 1. The spin expectation values for the electronic states near the Fermi level.
Figure 2. The electronic density of states (DOS) of GeTe monolayer.
Figure 3. The magnetic and superconducting properties of hole-doped bulk GeTe.
Figure 4. The superconductivity of bulk lead.
Figure 5. The superconductivity of electron-doped MoS$_2$ monolayer without/with magnetic field.
Figure 6. The superconductivity of Mn-doped GeTe monolayer.
Figure 7. The ferromagnetic Curie temperatures $T_c^{FM}$ of Mn-doped bulk GeTe.
Figure 8. The effect of magnetic anisotropy on the TSC phase.
Figure 9. The effect of GeTe film thickness on the TSC phase.
**Supplementary Table**
Table S1. The convergence of EPC λ.

**Supplementary Notes**

**Note 1. Determining the spin-state of Mn dopant**

Since two possible spin states of Mn atoms (high-spin state, S=5/2, and intermediate-spin state, S=3/2) exist under different conditions, it is necessary to determine the most favorable state in $Ge_{1-x}Mn_xTe$. The virtual crystal approximation (VCA) was employed to simulate Mn doping due to its computational efficiency. The direction of ferromagnetic order was set to along the (111) direction of bulk GeTe with Mn dopants, which is the easy magnetization direction for small $x$ [53]. The calculated Zeeman gaps of bulk $Ge_{1-x}Mn_xTe$ show similar trends with the experimental report [46] for both high- and intermediate-spin states (Supplementary Fig. 3c), but the former one appears to reproduce the experimental results better for $x<0.06$, indicating the Mn dopants are in $Mn^{2+}$ state. The slight deviation among them may be attributed to the antiferromagnetic coupling between the randomly distributed nearest-neighbor Mn dopants in experimental samples.

**Note 2. The convergence of EPC λ**

We have checked different **q/k**-point sampling meshes in the BZ of 0.1-hole-doped GeTe monolayer to ensure the convergence of electron-phonon coupling (EPC) calculation. To do so, we first check the convergence with respect to Gaussian broadening. The converged DOS at the Fermi level, $N_F \sim 1.4$ states/eV/primitive-cell (Supplementary Fig. 2), and the position of the Fermi level, characterized by the energy difference $\varepsilon_F \sim 272$ meV (Fig. 1b) between the Fermi level and the maximum of Ising bands, are used as the benchmarks. Using a Gaussian broadening of 0.005 Ry, $\varepsilon_F$ and $N_F$ are calculated to be 274 (273) meV and 1.42 (1.41) states/eV/primitive-cell on a 18×18×1 (16×16×1) **k**-point mesh, respectively, which well reproduce the corresponding data.

Then the **q**-point mesh used to calculate the dynamic matrix and phonon frequency are checked using the Gaussian broadening of 0.005 Ry. Supplementary Table 1 presents the EPC λ calculated with different combinations of **k**-point and **q**-point meshes. It is clear that λ starts to converge for the 8×8×1 **q**-point mesh. We finally choose λ = 1.39 obtained from using 36×36×1 **k**-point for EPC calculation, 18×18×1 **k**-point and **q**-point for phonon calculation in the main text.

**Supplementary Table 1. The convergence of EPC λ.** The calculated λ using different **k/q**-point mesh with the Gaussian broadening of 0.005 Ry. The finer/coarser **k**-point meshes are 36×36×1/18×18×1, 32×32×1/16×16×1, 36×36×1/18×18×1, 32×32×1/16×16×1, and 36×36×1/18×18×1, and the corresponding **q**-point meshes are listed in the table below.

| **q**-point mesh | **6×6×1** | **8×8×1** | **9×9×1** | **16×16×1** | **18×18×1** |
|---|---|---|---|---|---|
| λ | 1.85 | 1.48 | 1.42 | 1.35 | 1.39 |

## Note 3. Validating the newly developed first-principles approach

Note 3.1 Reproducing the superconductivity with time-reversal symmetry

We first confirm the reliability of our proposed first-principles method in predicting the superconducting gap and $T_c$ with time-reversal symmetry by reproducing the superconductivity of three representative conventional superconductors, i.e. bulk lead (Supplementary Fig. 4) [58], bulk GeTe (Supplementary Fig. 3d) [44], and MoS$_2$ monolayer (Supplementary Fig. 5) [59, 60]. The Debye temperature $\theta_D$ of lead, GeTe, and MoS$_2$ monolayer is about ~105, ~200, and ~300 K [61, 62], respectively. Bulk lead is a good metal with intrinsic superconductivity. Its absolute pairing strength $g_{nn}$ is calculated to be 2.4 from the EPC $\lambda$ ~1.3 [58], $N_F$~0.5 states/eV/primitive-cell (Supplementary Fig. 4a), and $\mu^* = 0.1$. The $T_c$ was then calculated from Eq. 3 as ~6.5 K (Supplementary Fig. 4b), which agrees well the experimental value of 7.2 K. The estimated $\Delta$~0.95 meV is also in good agreement with that calculated from the density functional theory for superconductors (SCDFT) [58], and slightly smaller than the experimental value (~1.3 meV). Secondly, the hole-doped bulk GeTe possesses comparable $\varepsilon_F$ (Supplementary Fig. 3b) and $\theta_D$, which tend to enhance the Coulomb repulsion $\mu^*$ by reducing $\ln(\varepsilon_F/\theta_D)$ in the $\mu^* = \dfrac{\mu_c}{1+\mu_c \ln(\varepsilon_F/\theta_D)}$ [56]. Assuming $\mu^*$=0.2 and taking $\lambda$=0.51 and $N_F$=0.632 states/eV/primitive-cell [64], we estimate the $g_{nn}$ value of bulk GeTe with 2.1×10$^{21}$ holes/cm$^3$ to be ~0.49. The self-consistently calculated superconducting gap and $T_c$ is 48.6 μeV and 330 mK (Supplementary Fig. 3d), respectively, consistent with the experimental results [44].

Especially, for MoS$_2$ monolayer with 1.2×10$^{14}$ electrons/cm$^{-2}$, theoretical calculations have consistently overestimated $T_c$ in comparison with the experiment. While by taking into account the electron-electron correlation that includes spin and charge fluctuations, one theoretical work has shown that the EPC is reduced from $\lambda$~1.32 to $\lambda$ ~0.72 by ~45.5% due to the renormalized electronic states [63], and the resulting $T_c$ is in good agreement with the experiment [60]. Correspondingly, we use the reduced $\lambda$~0.72 together with $\mu^*$=0.1 and $N_F$=2.14 states/eV/primitive-cell (Supplementary Fig. 5a) to formulate the gap equation, and the $g_{nn}$ is determined to be 0.29. Again, the calculated superconducting gap (Supplementary Fig. 5b) reproduces well the tunneling conductance measurement (2 meV) [59], and $T_c$ ~14 K agrees reasonably well with the experimental value ~11 K [60].

Note 3.2 Reproducing the superconductivity without time-reversal symmetry

In this part, we further demonstrate capability and reliability of our proposed first-principles method in predicting the critical magnetic field/doping-concentration by calculating the in-plane critical magnetic field $B_{c2}$ of MoS$_2$ monolayer in comparison with experiment [65]. It is worth noting that apparently this cannot be done by conventional first-principles method.

Specifically, we consider an electron doping concentration of $1.2 \times 10^{14}$ cm$^{-2}$, simulated by adding 0.1 electron per primitive cell into MoS$_2$ monolayer. At this doping concentration, our calculated superconducting gap (Supplementary Fig. 5b) without magnetic field reproduces very well the tunneling conductance measured gap of 2 meV [59]; the calculated T$_c$ ~14 K is only slightly higher than the experimental value ~11 K [60]. The electronic band structure is shown in Supplementary Fig. 5c. One clearly sees that the energy difference between the two valleys at the K point and between the K and Γ points is decreased, consistent with previous theoretical report. The MLWFs and the corresponding electronic Hamiltonian $H_{MLWFs}(\mathbf{k})$ were constructed by fitting the electronic band structure using the $d$ orbital of Mo atom as the initial guess for the unitary transformations.

The in-plane magnetic field was simulated by substituting the Pauli matrix $\sigma_z$ of Eq. 4 with $\sigma_x$. Then the electronic Hamiltonian of MoS$_2$ monolayer is re-written as:

$$H^x_{MLWFs}(\mathbf{k}) = B_x \sigma_x \otimes I_{\frac{N}{2} \times \frac{N}{2}} + H_{MLWFs}(\mathbf{k})$$

where $B_x$ is the in-plane Zeeman energy. Next, the new BdG Hamiltonian $H_{BdG}(\mathbf{k})$ was constructed through Eq. 1 and Eq. 2. The gap equation (Eq. 3) was solved self-consistently on a 200×200×1 **k**-points sampling in first BZ using the absolute pairing strength $g_{nn}$ of 0.29. The results for a series of in-plane Zeeman energy $B_x$, e.g. 0, 1, 2, 3, 4, 5, 6, 7, 8, 9 and 10 meV are shown in Supplementary Fig. 5b. One sees that the superconducting gap and T$_c$ are gradually suppressed with the increasing $B_x$. The $B_x$ can be transformed into magnetic field ($B$) in units of tesla (T) as $B = \mu_B B_x$, $\mu_B$ is the Bohr magneton. The in-plane upper critical magnetic field $B_{c2}$ for the MoS$_2$ monolayer is normalized by the Pauli paramagnetic limit (in units of tesla) $B_p = 1.86 T_c$. The $B_{c2}/B_p$ as a function of $T_c/T_{c,0}$ are shown in Supplementary Fig. 5d, which clearly indicates that our calculated critical magnetic field agree well with the experiments [65].

**Note 4. The effect of magnetic anisotropy on the TSC phase**

Given that the signature of Majorana fermions depends highly on the orientation of the magnetic field, it is necessary to discuss the effect of magnetic anisotropy on the TSC of Ge$_{1-x}$Mn$_x$Te monolayer. The electronic band structure of Ge$_{0.99}$Mn$_{0.01}$Te monolayer with the angle $\eta$ between the ferromagnetic ordering direction of Mn dopant and the normal direction of monolayer being set at 30° is shown in Supplementary Fig.

8a. One obvious difference from the case of $\eta = 0°$ (surface normal direction) is that the opened Dirac point in the Rashba bands moves away from the Γ point due to no-zero in-plane magnetic components. The Zeeman gaps ($\delta_z$) in the Ising and Rashba bands for different Mn concentrations with different magnetic directions are shown in Supplementary Fig. 8b and 8c, respectively. One can clearly see that $\delta_z$ decreases with the increasing $\eta$, which follows a relationship of $\delta_z(\eta) = \delta_z(0) \times \cos(\eta)$ perfectly. Notably, $\delta_z$ remains open for different angle $\eta$, indicating the robust TSC of $Ge_{1-x}Mn_xTe$ monolayer as long as $\eta \neq 90°$. For $\eta = 90°$, no Zeeman gap opens in either Ising or Rashba band due to a zero in-plane magnetic component at the Dirac point, so that the topological invariant is not well-defined.

To demonstrate the topological non-triviality of $Ge_{1-x}Mn_xTe$ monolayer for $\eta \neq 90°$, we take $\eta = 30°$ as an example by adding a Zeeman energy $B_x = 0.01\ eV$ and $B_z = 0.01\sqrt{3}\ eV$ into the electronic Hamiltonian $H_{MLWFs}(\mathbf{k})$:

$$H^{xz}_{MLWFs}(\mathbf{k}) = B_x \sigma_x \otimes I_{\frac{N}{2} \times \frac{N}{2}} + B_z \sigma_z \otimes I_{\frac{N}{2} \times \frac{N}{2}} + H_{MLWFs}(\mathbf{k})$$

A Zeeman gap of ~2 meV is opened in the Rashba bands, corresponding to $0.01 < x < 0.02$ (Supplementary Fig. 8c). Then the BdG Hamiltonian $H^{xz}_{BdG}(\mathbf{k})$ was constructed using Eq. 1 and Eq. 2 with the superconducting gap being set to 0.6 meV. By diagonalizing $H^{xz}_{BdG}(\mathbf{k})$ in the momentum space, we obtain the dispersion relation of superconducting quasi-particles (Supplementary Fig. 8d). The continuously opened superconducting gap enable us to calculated the total Berry curvature $\Omega = \sum_l \Omega^l(\mathbf{k}) = \sum_l \nabla \times A^l(\mathbf{k})$ for the states below the gap, where $A^l(\mathbf{k}) = i\langle u^l(\mathbf{k}) | \partial_\mathbf{k} u^l(\mathbf{k}) \rangle$ is Berry connection. Our calculated result (red lines in Supplementary Fig. 8d) shows that $\Omega$ is mainly located at the points associated with the Zeeman gap. Integrating $\Omega$ leads to the first Chern number $N_c = -1$, confirming the the existence of TSC states.

**Note 5. The effect of GeTe film thickness on the TSC phase**

Next, we briefly discuss the TSC of GeTe thin film with Te- and Ge-terminated polar surface (Supplementary Fig. 9a). The lattice constant increases with the increasing film thickness (Supplementary Fig. 9b), which will disentangle the Ising and Rashba bands starting from two layers (Supplementary Fig. 9c). Despite this, the TSC phase is still expected to form in GeTe thin film since the Rashba spin-splitting localized at the Te-terminated polar surface is well preserved. Doping 0.16 holes per primitive cell will move the Fermi level to the Dirac point of the two-layer GeTe film. One question is the stability of the polar surface, whose states will be shifted towards metallic nature (Supplementary Fig. 9d) due to the existence of surface dipole. Possible surface reconstructions need to be further explored experimentally and theoretically. We expect that surface reconstruction will not be strong enough to eliminate the Rashba spin-splitting, since the iconicity of Ge and Te is relatively week (compared with, let's say, oxides or nitrides) so that the surface dipole is weak. This is consistent with the observed possible signatures of TSC under external magnetic field in the GeTe thin film grown on Si(111) wafers [35]. Similarly, we believe that our predicted TSC in $Ge_{1-x}Mn_xTe$ monolayer is also feasible to be experimentally confirmed, albeit without the need of applying an external magnetic field.

**Supplementary Figures**

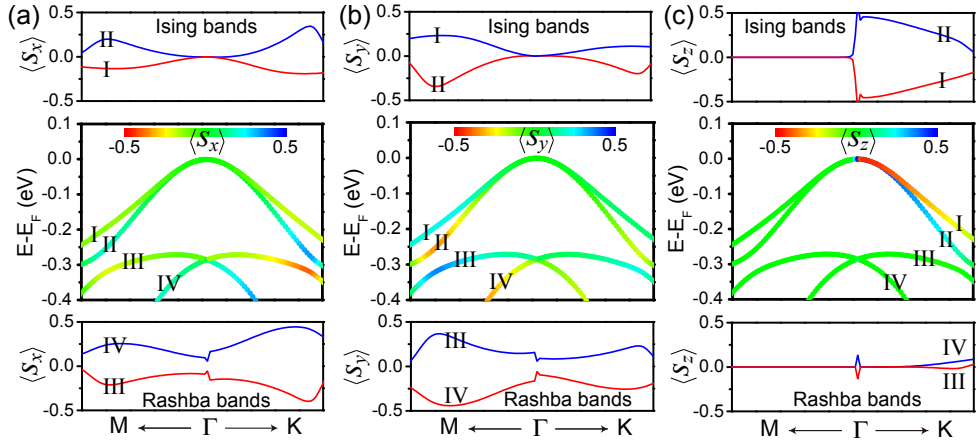

**Figure 1. The spin expectation values for the electronic states near the Fermi level.** (**a**) *x*-, (**b**) *y*-, and (**c**) *z*-components of the electron spin expectation value in the four valence bands near the Fermi level of our interest. Near the Γ point, the upper two bands (marked by I and II) are called as Ising bands and the lower ones (marked by III and IV) as Rashba bands, respectively.

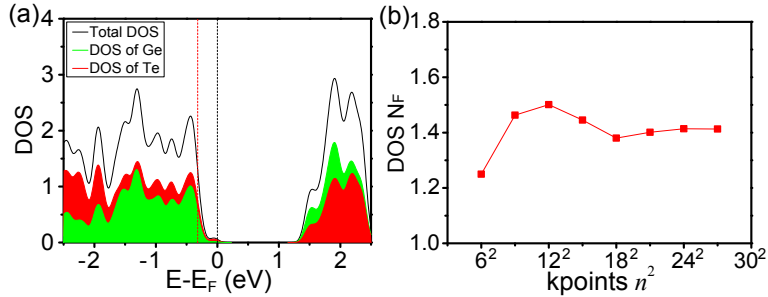

**Figure 2. The electronic density of states (DOS) of GeTe monolayer.** (**a**) The plot of the total DOS of GeTe monolayer. The partial DOS contributed by the *p*-orbitals of Ge and Te are represented by the green and red color, respectively. The black dashed line represents the Fermi level and the red dashed line is the Fermi level after doping 0.1 holes per primitive-cell. The unit of DOS is states/eV/primitive-cell. (**b**) The calculated DOS $N_F$ at the Fermi level of 0.1-holes-doped GeTe monolayer using different **k**-point sampling in the first BZ, which nearly converges to $N_F$ ~1.4 states/eV/primitive-cell starting from the 18×18×1 **k**-point mesh.

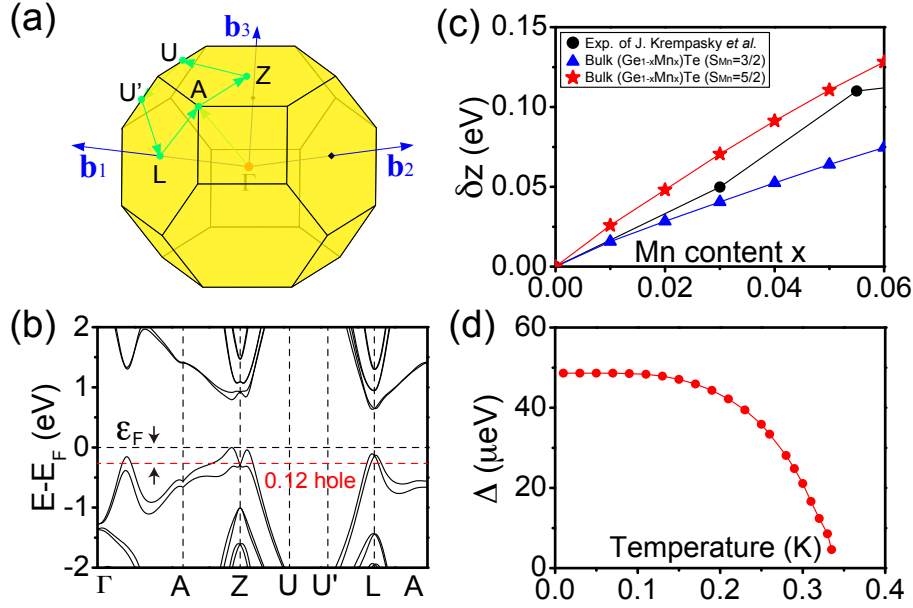

**Figure 3. The magnetic and superconducting properties of hole-doped bulk GeTe.** (**a**) The first BZ of bulk GeTe and the high-symmetry points used for calculating (**b**) band structure. The black horizontal dashed line in (**b**) represents the Fermi level and the red one is the Fermi level after doping 0.12 holes per primitive cell, corresponding to a hole doping concentrations of $\sim 2.1 \times 10^{21}$ holes/cm$^3$. (**c**) The comparison between the Zeeman gaps at the Z point reported by Krempasky *et al* (black dots) [46] and calculated by constraining the spin states of Mn dopants to the intermediate-spin (blue triangles) and high-spin (red stars) states based on VCA. (**d**) The temperature-dependent superconducting gap Δ of bulk GeTe with $\sim 2.1 \times 10^{21}$ holes/cm$^3$. The MLWFs were established by using the *p* orbital of Ge and Te atom as the initial guess for the unitary transformations and the superconducting gap was solved self-consistently on 100×100×100 **k**-points sampling in first BZ.

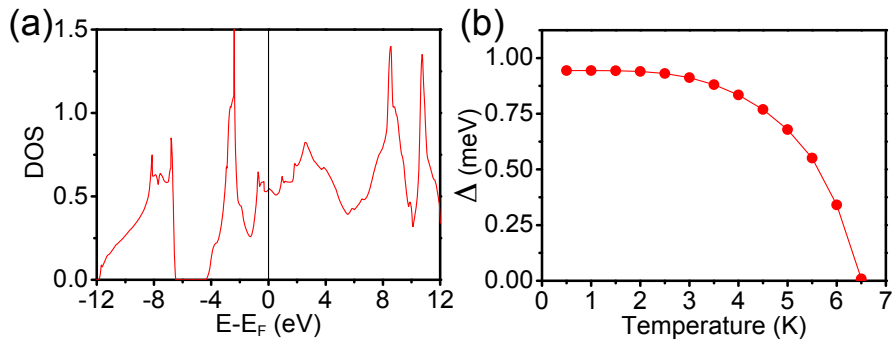

**Figure 4. The superconductivity of bulk lead.** (**a**) The total DOS of bulk lead. The unit is states/eV/primitive-cell. (**b**) The temperature-dependent superconducting gap Δ of bulk lead. The MLWFs were established by using the *p* orbital of Pd atom as the initial guess for the unitary transformations and the superconducting gap was solved self-consistently on 100×100×100 **k**-points sampling in first BZ.

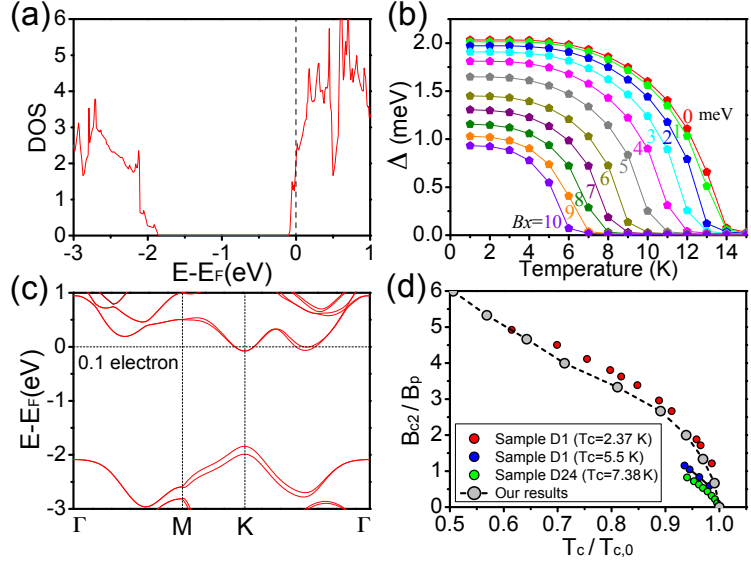

**Figure 5. The superconductivity of electron-doped MoS$_2$ monolayer without/with magnetic field.** (**a**) The total DOS of MoS$_2$ monolayer. The unit is states/eV/primitive-cell. (**b**) The temperature-dependent superconducting gap $\Delta$ of MoS$_2$ monolayer for a series of in-plane Zeeman energy $B_x$, e.g. 0, 1, 2, 3, 4, 5, 6, 7, 8, 9 and 10 meV. (**c**) The electronic band structure of MoS$_2$ monolayer doped with 0.1 electrons per primitive cell, corresponding to an electron doping concentration of $1.2 \times 10^{14}$ cm$^{-2}$. (**d**) The comparison between the calculated in-plane critical magnetic field of MoS$_2$ monolayer with the one measured by J. M. Lu et al. [65].

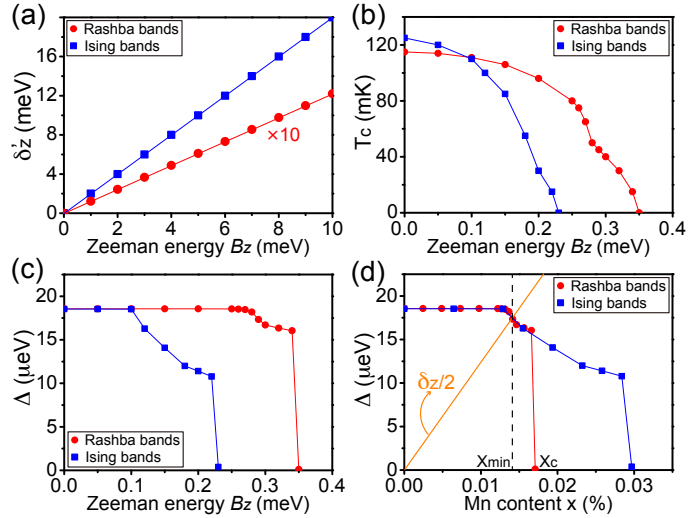

**Figure 6. The superconductivity of Mn-doped GeTe monolayer.** (**a**) The Zeeman gaps ($\delta'_z$) in the Ising/Rashba bands at the $\Gamma$ point versus Zeeman energy $B_z$. The $\delta'_z$ in the Rashba bands is magnified by 10 for clarity. (**b**) The transition temperature $T_c$ and (**c**) the superconducting gap of the Rashba/Ising bands versus Zeeman energy $B_z$. (**d**) The Mn concentration $x$-dependent superconducting gap of the Rashba bands (red dots) and Ising bands (blue squares). The orange line is drawn as $\delta_z/2 = 125 \times x$ meV, which is the half of the Zeeman gap opened at the Dirac point in the Rashba bands (Fig. 1d). The precondition of $\delta_z/2 > \Delta$ for TSC phase is fulfilled when $x > x_{\min} = 0.014\%$.

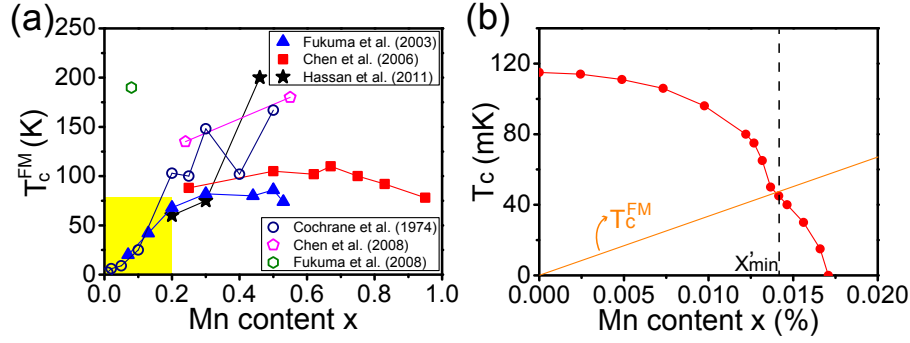

**Figure 7. The ferromagnetic Curie temperatures $T_c^{FM}$ of Mn-doped bulk GeTe.** (**a**) The $T_c^{FM}$ of $Ge_{1-x}Mn_xTe$ versus Mn concentration $x$, as reported by the experiments of Fukuma et al. (2003) [37], Chen et al. (2006) [38], Hassan et al. (2011) [39], Cochrane *et al.*(1974) [40], Chen et al. (2008) [41], and Fukuma et al. (2008) [42]. Overall, the $T_c^{FM}$ of hole-doped $Ge_{1-x}Mn_xTe$ (unfilled symbols) is higher than that without hole doping (filled symbols). The ferromagnetic Curie temperatures increase linearly with increasing Mn concentration up to $x$=0.2 (yellow back ground), and the linearly fitted equation is $T_c^{FM}(x)=333\times x$ K. (**b**) The comparison between the ferromagnetic Curie temperatures $T_c^{FM}$ and the Mn content $x$-dependent superconducting translation temperature $T_c$ of the Rashba bands, which cross over at $x'_{min}$=0.014%.

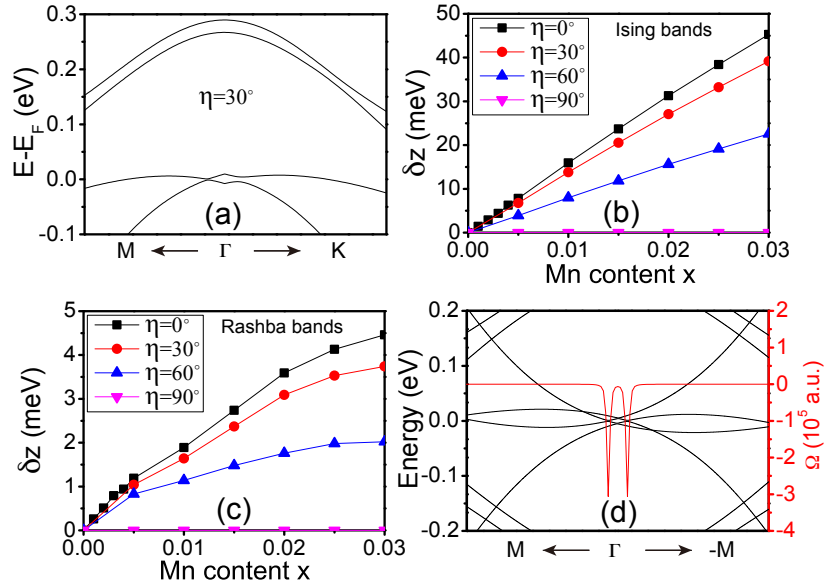

**Figure 8. The effect of magnetic anisotropy on the TSC phase.** (**a**) The band structure of $Ge_{1-x}Mn_xTe$ monolayer with $\eta$=30° and $x$=0.01. The $\eta$ is the angle between the ferromagnetization direction and the surface normal direction of the monolayer. The in-plane component of magnetization moves the Dirac point of the Rashba bands away from the Γ point. The Zeeman gaps ($\delta_z$) in (**b**) the Ising and (**c**) the Rashba bands versus the Mn concentration $x$, are calculated from first-principles under VCA. The angle $\eta$ is varied to reveal the effect of magnetic anisotropy on the Zeeman gaps. (**d**) The dispersion relation (black curves) of superconducting quasi-particles calculated from the first-principles BdG Hamiltonian. The red lines denote the Berry curvatures for the states below the superconducting gap.

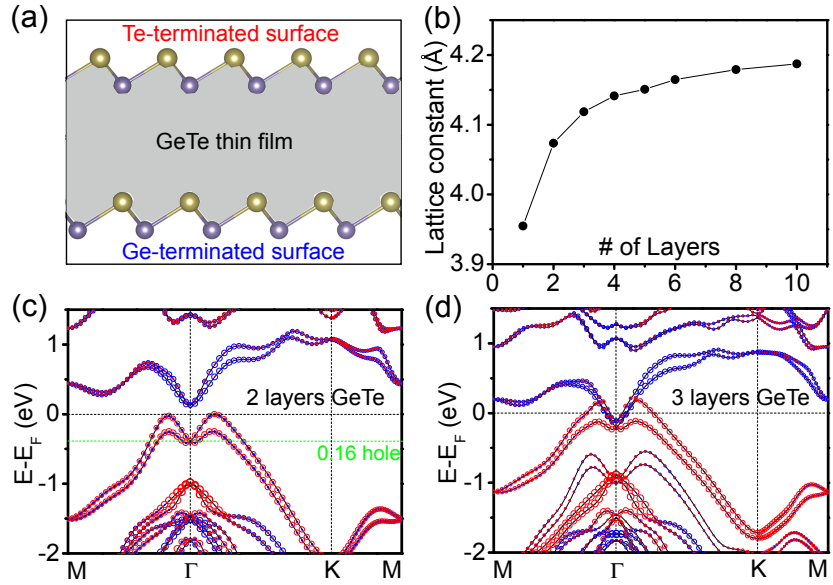

**Figure 9. The effect of GeTe film thickness on the TSC phase.** (**a**) The schematic diagram of GeTe thin film with Te- and Ge-terminated polar surface. (**b**) The variation of the in-plane lattice constant of GeTe thin film versus the number of layers. The electronic band structure of (**c**) the two-layer and (**d**) three-layer GeTe thin film. The size of red and blue circles is drawn proportional to the contributions from the Te- and Ge-terminated surface layer, respectively.